# Polarization sensitive laser intensity inside femtosecond filament in air

Xuan Zhang[1,2], Tie-Jun Wang[1,2,*], Hao Guo[1,2], Na Chen[1], Lei Lin[1,2], Lingang Zhang[1], Haiyi Sun[1], Jun Liu[1,2], Jiansheng Liu[3], Baifei Shen[3], Ruxin Li[1,2,*], and Zhizhan Xu[1,2]

[1] *State Key Laboratory of High Field Laser Physics, Shanghai Institute of Optics and Fine Mechanics, Chinese Academy of Sciences, Shanghai,201800, China*

[2] *Center of Materials Science and Optoelectronics Engineering, University of Chinese Academy of Sciences, Beijing, 100049, China*

[3] *Department of Physics, Shanghai Normal University, Shanghai 200234, China*

*Corresponding authors : tiejunwang@siom.ac.cn (TJW), ruxinli@mail.siom.ac.cn (RXL)*

## Abstract

Polarization dependent on clamping intensity inside femtosecond filament was experimentally measured in air. By tuning the laser pulse ellipse from linear polarization to circular polarization, the measured clamping intensity inside laser filament is gradually increased up to 1.7 times. The experimental results are in good agreement with the simulation results by solving the extended nonlinear Schrödinger equation (NLSE). The polarization sensitive clamping intensity inside filaments offers an important factor towards fully understanding the polarization related phenomenon observed so far.

Recently, there has been an ascending interest in a large variety of promising applications of the femtosecond laser filamentation [1-8] ranging from supercontinuum generation [9, 10], THz radiation [6, 7], air lasing [11, 12], guiding corona discharges [13, 14] to induce ionic wind [15] and etc. Under such enthusiasm, there are several special phenomena observed under specific laser polarization state for filamentation. For instance, the circularly polarized beam is less effectively inducing multiple filament [16], a strong backward stimulated radiation is observed from neutron nitrogen at 337nm only under circular polarization [17], as well as the efficiency of terahertz emission is enormously enhanced under the circularized polarized laser versus linearly polarized one [18]. Besides, the filament-induced supercontinuum generation is enormously enhanced with an elliptical polarization [19]. Few numerical simulations have been made to predict the laser polarization related filamentation process [20-22]. Kolesik et al [20] reported that the circularly polarized pulses created lower plasma densities as compared with the linearly polarized pulses which is contradicted with the experimental observations [23]. While Panov and co-workers' simulation [21] indicated a higher clamped intensity under circular polarization than under linear polarization. In 2013, the same group developed a theoretical model of femtosecond filamentation [22] which includes high-order Kerr effect and ignores the effect of plasma instead. Their simulation results show that the maximum filament laser intensity tends to the intensity of linearly polarized one when laser ellipticity $\varepsilon_0 \lesssim 0.2$; the maximum filament laser intensity tends to the intensity of circularly polarized one when laser ellipticity $\varepsilon_0 \gtrsim 0.3$; while the maximum filament laser intensity has a step-like increase when laser ellipticitty is in between 0.2 and 0.3. By tuning laser polarization, the control of femtosecond filamentation has been well demonstrated experimentally [24].

However, the physical picture of the polarization dependence on the filamentation behavior has not been elaborately illustrated. One general interpretation is the influence on the accelerating process of the free electron driven by the different polarization [25]. Circularly polarized laser filament favors the free electron acceleration [26]. Apparently, the effect of polarization on laser intensity (or strength of laser electric field) inside the filament is overlooked. The measurement of polarization dependent laser intensities inside a filament would benefit the understanding of laser polarization sensitive phenomenon which is still absent so far.

The direct measurement of filament intensity is nontrivial due to the high laser intensity induced material damage. To avoid damage of the detective devices, the intensity of the filament is generally obtained through indirect approaches [23,27]. By firstly measuring the plasma density through diffractometry [28], then the intensity could be calculated through a semi-experience model of molecules' tunnel ionization [27]. Recently, Mitryukovskiy et al [29] reported on a direct measurement of "burning hole". By inserting a metallic foil inside the filament, laser filament punched a pinhole with a diameter of the filament. Then the intensity inside the filament was obtained from the trivial measurements of filtered out pulse energy and duration of the filament pulse [29]. Following the method, Li et al [30] successfully measured the clamped intensity inside a filament in a flame.

In this paper, we present an experimental study of the polarization dependent on laser intensity inside an air filament by employing the "burning hole" approach. The experimental results show filament intensity gradually increases when the laser polarization ellipse tunes from linear to circular. Surprisingly the circularly polarized filament intensity is 1.7 times higher than the linearly polarized one. Then a numerical simulation based on Nonlinear Schrodinger equation (NLSE) by split-step Crank-Nicolson method is performed. The simulation results nicely confirm the experimental observations. The results in the work indicate that the polarization effect on filament intensity has to be considered in order to fully understand the polarization sensitive phenomena of the filament applications.

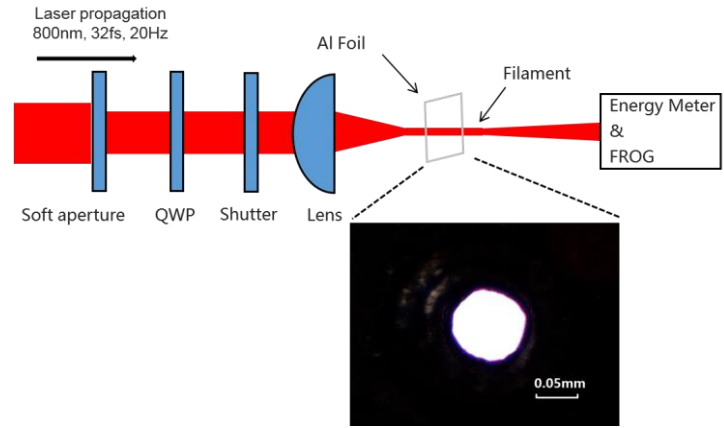

Fig. 1 Experimental Setup

The experimental layout is arranged as shown in Fig.1. The input laser pulses are generated from a Ti: sapphire chirped pulse amplification (CPA) laser system with central wavelength at 800 nm, 20 Hz repetition frequency and 32 fs pulse duration. A soft aperture [31] was used to get rid of any diffraction effect on the laser pulse and obtain an input laser beam with a clean profile. Then the initial linear polarization of the laser was tuned by a quarter-wave plate (QWP) to provide the laser beam with different initial ellipticities. A mechanical shutter was employed to control the laser pulse number passing through it. The input laser beam was focused by a plano-convex lens with 30cm focal length. Since laser intensity is clamped inside filament even with the increase of laser energy [1], the input laser energy per pulse used in the experiment ranges from few hundred μJ to 1.5 mJ with emphasis at the single filament region. Thus, with the help of the self-focusing and the defocusing effect of the plasma produced by ionization process, the laser filament presents near the geometrical focus of the focal lens. An aluminum (Al) foil with uniform thickness of around 20μm is strictly perpendicularly placed with respect to the laser propagation direction. All the Al foils used in this paper were placed slightly (0.5cm) before the lens geometrical focus to keep the Al foil at the intense region of filament. The transmitted filament energy and pulse duration were recorded by an energy meter (Coherent FieldMaxII) and a homemade frequency-resolved optical gating device (FROG), respectively.

When the incident laser power overtakes the critical power for self-focusing, the successive laser pulses acts as a positive lens tending to focus the beam on itself. The combination work of diffraction, dispersion and ensuing ionization of the air molecules acts against self-focusing process. Ultimately, a laser filament forms in air, which restricts amount of energy within a comparative scale and maintain this condition over the order of Rayleigh distance in the propagation direction. Due to its high intensity, the laser filament can be used as a light drill to ablate a pinhole on the Al foil. Since the spatial mode inside laser filament core is approximately fundamental mode [7], the transverse intensity distribution of filament core is nearly Gaussian type. The filament drilled pinhole's diameter changes as a function of the input laser pulses number due to accumulation effect. The result is shown in fig. 2 at 1mJ incident per pulse energy with linear polarization. The pinhole diameter verse input pulses number gradually rises from ~60μm after 20 shots to a stable value of ~100μm after 120 shots. When it comes to the stable state, the filament sufficiently destroys the Al foil and passes through the pinhole. The energy reservoir around the filament is not intense enough to break the metallic foil even if the shots number is comparatively large. Considering the ablation threshold of the Al foil, the real diameter of filament core is approximately 110-120μm which agrees with the experimental observation in other works at similar conditions [2]. In the experiment, a reference pinhole with diameter around 60μm formed after 20 shots of successive laser pulses was chosen for the following intensity measurements which is close to the filament diameter at FWHM [32]. The experimental procedure is followed: A reference pinhole of ~60μm on the Al foil was prepared first by using 20 shots of successive laser filament with 1mJ per pulse. Then the reference pinhole was carefully placed at the same position slightly before geometrical focus. By controlling the shutter, only a single shot is released to form filament in air for measurement. By tuning the QWP, the polarization of filamenting pulse is working at a fixed ellipse. Then the transmitted filament energy is measured by the energy meter sensor. The transmitted laser pulse is captured by the FROG device for pulse duration measurement. The size of the pinhole on the Al foil was evaluated by a transmissive microscope. On account of the definition of the intensity, the intensity inside a filament can be directly calculated by the pinhole's area, the transmitted energy and laser duration after filamentation. To eliminate the unpredictable fluctuation of the laser pulse, we repeat above-mentioned experiment process by 6 times in each case to reduce measurement errors. Then above procedure was repeated at different polarization ellipse.

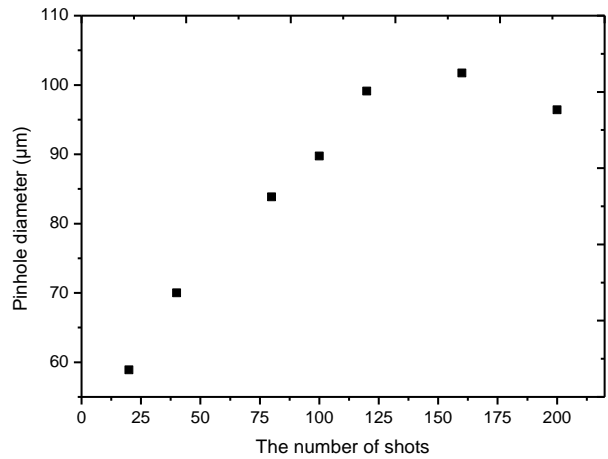

Fig. 2 The diameters of laser filament drilled Al foil pinholes as a function of the number of the laser shots

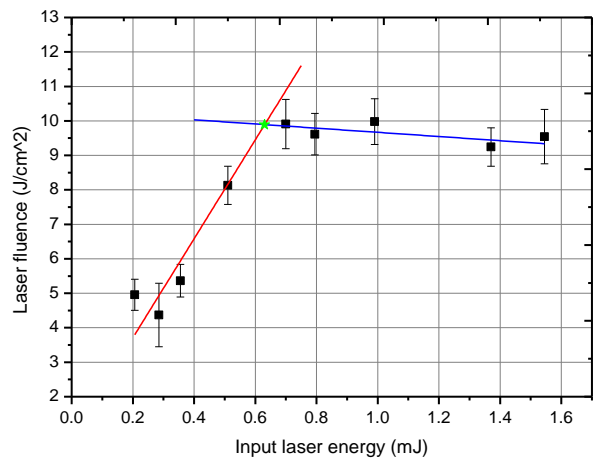

Fig. 3 Measured laser fluence as a function of the input laser energy at specific circular polarization (ellipticity of 1)

Fig. 3 shows the typical laser fluence inside ~FWHM diameter (60μm-diameter) filament core as the function of the input laser energy at a given initial ellipticity of 1 (circular polarization). The laser fluence linearly grows rapidly before reaching the saturated value around 10 $J/cm^2$. When the input laser energy is < ~0.6mJ, the laser fluence linearly increases with the increase of input pulse energy. This clearly indicates the linear stage before filamentation. More energy is confined in the certain beam area which is proportional to the input pulse energy. A clear saturation of the laser fluence is observed when the input laser energy is > ~0.6mJ which is in a good agreement with ref. [29]. The laser fluence stays almost at a constant value of ~10 $J/cm^2$ The saturation (clamping) of the laser fluence indicates the occurrence of filamentation. [7] The laser fluence inside filament core stays almost the constant value due to the counterbalance between self-focusing and plasma defocusing. The green star in fig. 3 marks the intersection point of these two fitting lines at 9.89 J/cm², 0.63mJ which is the measured filament fluence at the initial ellipticity of 1. By rotating the QWP, the incident laser fields with different initial ellipticity for filamentation were investigated. According to these intersection data, the polarization dependence of the filament fluence is presented in fig. 4. With the initial ellipticity tuned from 1 to 0, as polarization changes from circular polarization to linear polarization, the laser fluence inside the filament core decreases from $13.8 \, J/cm^2$ to $9.89 \, J/cm^2$. The ratio of laser fluence evaluated at initial circular polarization to linear polarization is about 1.4.

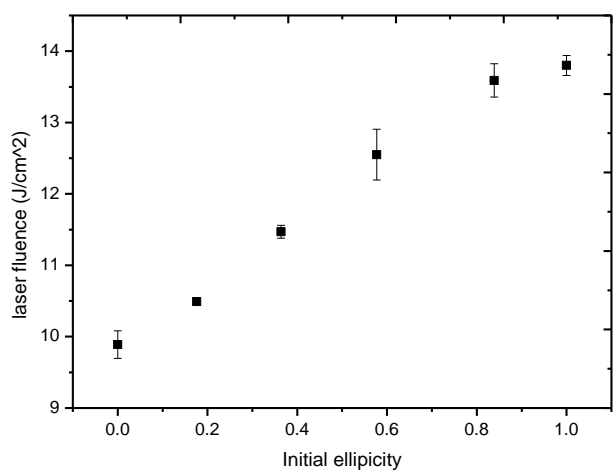

Fig. 4 Clamping laser fluence inside filament core versus different initial laser polarization ellipse

In order to calculate the intensity, the pulse duration after filamentation is obtain by a FROG device. The pulse duration before the filamentation is 32fs. However, after filamentation the retrieved temporal duration of initial circular polarization pulses and linear polarization are stretched to 60fs and 74fs, respectively, although the spectra measured after filamentation did not change in our experimental conditions. The apparently broaden of the pulse duration after filamentation could stem from filamentation induced angular dispersion together with the temporal dispersion [33]. According to the pulse durations obtained, the measured laser intensities inside filament core is $2.27 \times 10^{14} \, W/cm^2$ for circularly polarized input laser. For linear polarization, the intensity is about $1.33 \times 10^{14} \, W/cm^2$, which is 1.7 times smaller than the circular polarization one.

In order to verify the experimental results as well as get an insight into the physical essence of the polarization dependence of filament intensity, a numerical simulation by solving the NSLE is performed based on split-step Crank-Nicolson method [34]. Assuming an axially symmetrical laser beam with Gaussian profile in both time and space domain which propagates along the z axis, the input laser beam can be depicted as

$$\varepsilon(r, t, z) = A_0 * \exp\left(-\frac{r^2}{r_0^2} - i\frac{k_0 r^2}{2Fl} - \frac{t^2}{t_p^2}\right). \tag{1}$$

Here, $A_0$ is amplitude of incident laser electronic field, $r = (x^2 + y^2)^{0.5}$, $r_0 = 6mm$ is the waist of initial beam, $k$ is the central wave vector, Fl = 0.3m is the lens focal length and $t_p = 27fs$ (corresponding to the initial pulse full width at half maximum (FWHM) duration $t_{FWHM} = t_p\sqrt{2\log 2} = 32fs$)[35]. Also, the incident laser beam can be represented into two components $\varepsilon = \varepsilon^+ + i*\varepsilon^-$ which describes the orthogonal polarized part of the initial polarized beam with a fixed phase difference of $0.5\pi$. In order to describe the arbitral initial polarization state, a parameter of ellipticity $\theta = |A^-|/|A^+|$ is introduced, where the $|A^+|$, $|A^-|$ represent the amplitude of $\varepsilon^+$, $\varepsilon^-$ respectively. Thus, to change the value of $\theta$ is to change the initial polarization. Note that $\theta = 0$ and $1$ indicates the input laser polarization are linearly and circularly polarized, respectively. To keep a constant initial amplitude under defferent initial polarization state an amplitude index of $F^\pm$, with $A^\pm = F^\pm * A_0$ and $F^{+2} + F^{-2} = 2$ is introduced.

In the meanwhile, the propagation process can be described in a straight forward extension of scalar version of NLSE [34] as

$$\frac{\partial \varepsilon^\pm}{\partial z} = \frac{i}{2k}\nabla_\perp^2 \varepsilon^\pm - \frac{ik''}{2}\frac{\partial^2 \varepsilon^\pm}{\partial t^2} - \frac{\beta^{(K)}}{2}|\varepsilon|^{(2K-2)} + i\frac{2\omega}{3c}fn_2\left(|\varepsilon^\pm|^2 + 2|\varepsilon^\mp|^2\right) - \frac{\sigma}{2}(1+i\omega t)\rho\varepsilon^\pm. \quad (2)$$

The terms on the right hand indicate the diffraction effect, normal group velocity dispersion (GVD), multiphoton ionization (MPI), instantaneous cubic nonlinearity and the plasma effects. Here $\omega$ is the frequency of the input beam, $k'' = \partial^2 k/\partial \omega^2$, $\beta^{(K)}$ is the coefficient of MPI, $K$ is the minimum photon number for MPI, $c$ is the speed of light, $f$ is the ratio of self- and cross- nonlinearity effect, $\sigma$ is the collision cross section of inverse bremsstrahlung and $\rho$ is the density of laser induced plasma. To complete the simulation, the plasma density follows the evolution equation as

$$\frac{\partial \rho}{\partial t} = \frac{\sigma}{E_g}\rho|\varepsilon|^2 + \frac{\beta^{(K)}|\varepsilon|^{2K}}{K\hbar\omega} - \alpha\rho^2. \quad (3)$$

The first term on the right hand represent the avalanche ionization and $E_g$ is the ionization energy. The second term on the right hand is MPI. The last one corresponds to recombination process of the electrons with a rate of $\alpha$. The values of these parameters employed in our simulation are adopted from ref. 36 & 37.

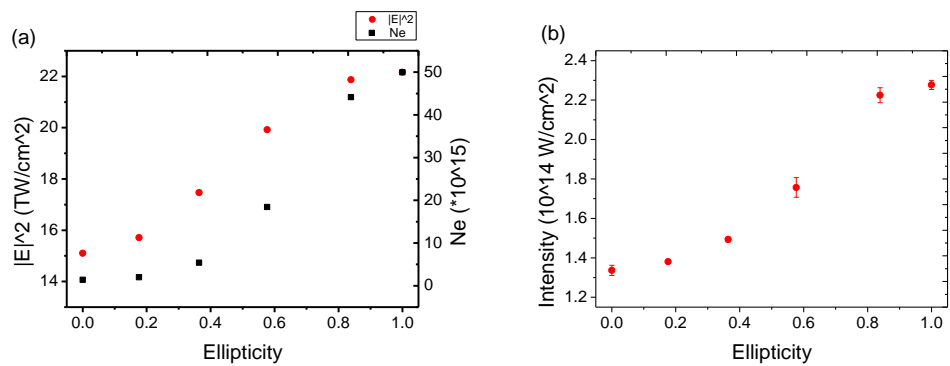

Fig. 5 filament intensity versus the initial laser polarization ellipticitty: (a) simulation results of $|E|^2$ and electron density $N_e$, (b) clamping intensity insider filament core measured in the experiment. The ellipticity of 0 and 1 in (a), (b) are for the linear polarization and circular polarization, respectively.

Since laser intensity I is proportional to the square modulus of the electric field $|E|^2$, the square modulus of the electronical field is used to represents the intensity, for instance, $I \propto |\varepsilon^2| = |(\varepsilon^+)^2| + |(\varepsilon^-)^2|$ in the simulation. The critical power for self-focusing collapse under linearly polarization is $P_{cr} = 3.18GW$ ($P_{cr} = \lambda_0^2/2\pi n_2$) [36]. In simulation, the input laser power $P_{in} = 4.5P_{cr}$, corresponding to the input laser energy around 1mJ. The simulated peak intensity inside the filament gradually increases as the initial semi-axis ratio rising from 0 to 1, implying with the initial polarization state changing from linear polarization to circular polarization as shown in Fig. 5(a). Compared with the simulation results in Ref. [22], no intensity jump at specific ellipticity region is observed in our simulation. The simulated results in this work nicely agrees with the experimental observation in Fig. 5(b). Clearly, the simulated peak intensity in CP is ~1.46 times higher than in LP which is very close to the experimental results of ~1.7.

In the meantime, the calculated plasma density induced by ionization process also varies under different input laser initial ellipticity as shown in Fig.5(a). As θ changes from 0 (LP) to 1 (CP), plasma density increases from nearly $1 \times 10^{15}$ to $5 \times 10^{16}$, which corresponds to the incline trend of peak intensity. Conceivably, when the intensity is clamped due to the balance between self-focusing and laser induced plasma, the change of index of refraction would be commensurate. Given the higher order Kerr effects are generally neglected in the filamentation process, such restrict relationship can be roughly described as $\Delta n_{Kerr} \approx n_{plasma}$ [1]. In air, it can be expressed with the assumption that the $\rho(I)$ mainly generated by MPI, as $n_2 I \approx \rho_{MPI}(I)/2\rho_c$, $\rho_c = \epsilon_0 m_e \omega_0^2 / q_e^2$ denotes the critical density of the plasma with $\epsilon_0$, $q_e$, $m_e$ being the permittivity of vacuum, elementary charge and the electron mass, respectively. Due to the higher ionization ratio and higher critical power for self-focusing at the laser polarization of CP [22], more energy confines inside the core before filamentation resulting in the higher intensity inside filament. Thus, the higher intensity has to be balanced by the higher generated plasma density during filamentation. Then the laser intensity inside filament core is clamped at a constant value. In our simulation, circularly polarized laser induces denser plasma than the linearly polarized one. This, in turn, support the calculated result of simulated peak intensity variation trend as well as the experimental results.

In conclusion, we systematically investigate the input laser polarization effect on the intensity inside an air filament. Through a simple method of the filament burning hole on Al foils, the laser polarization dependent clamping intensity inside a filament was directly measured in air. As the incident laser is circularly polarized for filamentation, the measured laser intensity clamped inside the filament is ~1.7 times higher than the one induced by linearly polarized laser. The clamping intensity inside filament gradually decreases when the input laser polarization is tuned from CP to LP. A simulation based on the direct extended NLSE is performed. The simulated results of laser intensity and plasma density support the experimental observation in a good agreement. When filamenting laser polarization tunes from LP to CP, more energy confines inside the filament core resulting in the higher intensity. The results in this work not only clarify the disputes on the prediction of filament clamping intensity [6], but also benefit the experiments. Although the experiments and simulations are performed in air, the results may apply for other filamentation medias. The polarization sensitive clamping intensity results inside a filament has to be considered towards understanding the mechanisms of polarization dependent phenomenon observed [23,24].

## Acknowledgements

This work was in part supported by the Strategic Priority Research Program of the Chinese Academy of Sciences with Grant No. XDB160104, the International Partnership Program of Chinese Academy of Sciences with Grant No. 181231KYSB20160045.